\documentclass[prb,twocolumn,aps,superscriptaddress,showpacs,floatfix]{revtex4-2}
\usepackage{amsbsy,amssymb,amsmath,bm}
\usepackage{graphicx,color,epsfig,rotate}
\usepackage{xspace,units}
\usepackage{textcomp}		%for upright \textmu
\usepackage{epstopdf}
\usepackage{tabularx}

\usepackage{multirow}		
\usepackage{dcolumn}							
\newcolumntype{d}[1]{D{.}{.}{#1}}
 
\newcommand{\Gd}{Gd(NO$_3$)$_3$~}
\newcommand{\Nd}{\mbox{Nd-Fe-B}\:}

\begin{document}
\title{Neutron Imaging of Paramagnetic Ions: \\ Electrosorption by Carbon Aerogels and Macroscopic Magnetic Forces}

\author{Tim A. Butcher}
\email{tbutcher@tcd.ie}
\affiliation{School of Physics and CRANN, Trinity College, Dublin 2, Ireland}

\author{Lucy Prendeville}
\affiliation{School of Physics and CRANN, Trinity College, Dublin 2, Ireland}

\author{Aran Rafferty}
\affiliation{AMBER Centre and School of Chemistry, Trinity College, Dublin 2, Ireland}

\author{Pavel Trtik}
\affiliation{Laboratory for Neutron Scattering and Imaging, Paul Scherrer Institut, CH-5232 Villigen, Switzerland}

\author{Pierre Boillat}
\affiliation{Laboratory for Neutron Scattering and Imaging, Paul Scherrer Institut, CH-5232 Villigen, Switzerland}
\affiliation{Electrochemistry Laboratory, Paul Scherrer Institut, CH-5232 Villigen, Switzerland}

\author{J. M. D. Coey}
\affiliation{School of Physics and CRANN, Trinity College, Dublin 2, Ireland}

%\date{\today}

\begin{abstract}

\noindent  The electrosorption of Gd$^{3+}$ ions from aqueous 70\,mM \Gd solution in monolithic carbon aerogel electrodes was recorded by dynamic neutron imaging. The aerogels have a bimodal pore size distribution consisting of macropores and mesopores centered at 115\,nm and 15\,nm, respectively. After the uptake of Gd$^{3+}$ ions by the negatively charged surface of the porous structure, an inhomogeneous magnetic field was applied to the system of discharging electrodes. This led to a convective flow and confinement of \Gd solution in the magnetic field gradient. Thus, a way to desalt and capture paramagnetic ions from an initially homogeneous solution is established.

\end{abstract}

%\pacs{71.18.+y, 71.27.+a, 74.70.Tx}

\maketitle

%----------------------------------------------------------------------------------------------------------------------------------------------------------------------------------------------
\section{Introduction}%------------------------------------------------------------------------------------------------------------------------------------------------------------------------
%----------------------------------------------------------------------------------------------------------------------------------------------------------------------------------------------
Desalination is the process of removing dissolved ions from water. The minimum energy required to separate 1\,M of solvated ions from bulk aqueous solution is related to the change in chemical potential $\Delta E_{\mathrm{\mathrm{min}}} = - R T \, \mathrm{ln}(x_{i}) = 10\,\mathrm{kJ}\,\mathrm{mol}^{-1}$ (\mbox{$R$: gas constant}, $T$: room temperature and $x_i$: mole fraction of the ions). This is a high energy barrier, which must be overcome in any desalination process.  
The magnetism of paramagnetic ions, such as those of transition or rare earth metals, is preserved  when the corresponding salt is dissolved in water. When an inhomogeneous magnetic field with gradient ${\nabla B}$ is applied to such solutions, it exerts the magnetic field gradient force \cite{remark_kelvin}:

\begin{equation}
\mathbf{F}_{\nabla B} = \frac{\chi}{\mu_0 } (\mathbf{B}\cdot \nabla) \mathbf{B}.
%\mathbf{F}_{\nabla B} = \frac{\chi}{2\mu_0 } \nabla \mathbf{B}^2.
\end{equation}

\begin{figure}
	\centering
	\includegraphics[width=0.99\columnwidth]{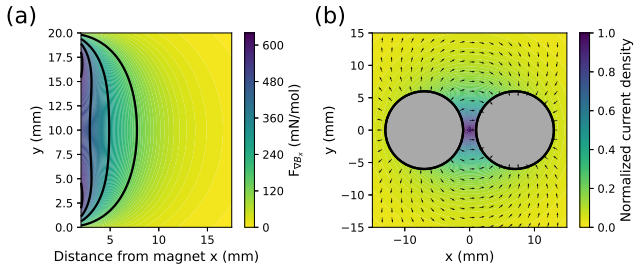}
	\caption{(a) Magnetic field gradient force distribution for Gd$^{3+}$ ions in the field of a uniformly magnetized 20\,mm \Nd cube. The magnetic charge model was used \cite{furlani_2001}. (b) Normalized current distribution between two cylindrical electrodes of 12\,mm diameter in a liquid with uniform conductivity. The separation between the electrodes is 2\,mm.}
	\label{fig:forces}
\end{figure}

This is a force density with the magnetic susceptibility of the solution $\chi$ and the permeability of free space $\mu_0$ (see Fig.~\ref{fig:forces}(a) for an example with Gd$^{3+}$ ions in the stray field of a permanent neodymium-iron-boron (Nd-Fe-B) magnet). It allows magnetic levitation of submerged objects \cite{andres_1966, dunne_2007} or the inhibition of density-difference driven convection \cite{coey_2009, butcher_2020}. The latter is of particular interest, since it follows that paramagnetic liquids can be manipulated in non-magnetic miscible liquids. The magnetically modified concentration profile is eventually homogenized by diffusion. This is readily understandable when comparing the aforementioned chemical potential, the derivative of which governs diffusion, and the magnetic energy given by $E_\mathrm{mag} = \tfrac{\chi_m}{2\mu_0} B^2$ ($\approx 130$\,mJ\,mol$^{-1}$ for Gd$^{3+}$ with molar susceptibility $\chi_m = 330\times 10^{-9}$\,m$^3$\,mol$^{-1}$ \cite{coey_2009} at $B=1$\,T). Considering these values, a magnetic separation of ions could be deemed an Icarian endeavour. Nevertheless, attempts of rare earth ion separation in a magnetically modified Clusius-Dickel thermal diffusion column \cite{muller_1988,brandon_2021} were reported by Ida and \mbox{Walter Noddack} in the 1950s \cite{noddack_1952, noddack_1955, noddack_1958}. After a 50~year hiatus, interest in the effects of magnetic field gradients on ionic solutions has been revived \cite{coey_2009, eckert_2012, eckert_2014, rodrigues_2017, eckert_2017, rodrigues_2019, lei_2020, lei_2021, ji_2016, franczak_2016, kolczyk_2016,rodrigues_2018, kolczyk_2019, higgins_2020, butcher_2020, fritzsche_2020}. Magneto-convection only arises when the magnetic field gradient force is rotational, which necessitates a susceptibility gradient and magnetic field gradient that are non-parallel \cite{mutschke_2010}. Interferometric measurements generated evidence of a magnetic ion enrichment around the surface of an evaporating solution containing a single paramagnetic species  \cite{eckert_2012,eckert_2014,rodrigues_2017,eckert_2017,rodrigues_2019, lei_2020, lei_2021}. The inhomogeneity in these systems is caused by the input of thermal energy, which is the driving force for the concentration change and essentially a distillation process. The energy consumption by water evaporation is 40.65\,kJ\,mol$^{-1}$, unquestionably surpassing the necessary threshold set by the chemical potential. %2.5\,kJ\,g$^{-1}$.

A further driving force on ions is generated by gradients of an electric potential $\frac{\textrm{d} \Phi}{\textrm{d} r}$ and the subsequent movement is known as migration. The corresponding expression for the force is \mbox{$F_{el} = - F Z \frac{\textrm{d} \Phi}{\textrm{d} r}$}, with the Faraday constant (\mbox{$F = 96\,485\,$C\,mol$^{-1}$}) and the valence charge of the ion $Z$. An electric double layer forms at the interface between a charged electrode and an electrolytic solution. This region is in the order of 3\,nm with a potential gradient of $\frac{\textrm{d} \Phi}{\textrm{d} r} = 10^7\,$V\,m$^{-1}$. Hence, the resulting forces, ${\sim} 10^{12}$\,N\,mol$^{-1}$, are sufficiently gigantic for charge separation and a mass transport-limited diffusion layer is formed through which the ions travel. Thus, a current with a diffusion and migration component flows. The current distribution in a cell with two cylindrical electrodes is shown in Fig.~\ref{fig:forces}(b).

An approach for the continuous removal of ions from water is capacitive deionization \cite{porada_2013,suss_2015}. This relies on the immobilization of ions in the electric double layer of porous electrodes with enormous surface areas. Once the pores of the electrode are filled, the fully charged electrodes can be discharged and the ions are released into the surrounding liquid. Then, the cycle begins anew. This cyclic process in which the ions are not transformed into a solid makes the procedure more attractive for hard to plate ions, such as rare earths. What is more, capacitive deionization is far more energy efficient than a desalination based on evaporation. %surface areas (- \,m$^2$\,g$^{-1}$)

Although magnetic forces pale in comparison with those of electric nature, previous studies have shown that electrodeposits from solutions containing paramagnetic ions can be structured on electrodes with magnetic field gradients \cite{tschulik_2009, tschulik_2010, mutschke_2010, tschulik_2011, tschulik_2012, dunne_2011, dunne_2012}. In these depositions, the paramagnetic ions in solution are converted at the working electrode, causing a concentration gradient and convective flow which the magnetic field gradient drastically alters. In comparison, the influence of a magnetic field gradient on an electrochemical cell with desalinating porous electrodes remains uncharted territory.

We report a neutron imaging study of the capacitive deionization of a paramagnetic \Gd solution with a magnetic field gradient imposed during the discharge process. Unlike methods such as small-angle neutron scattering that provide information on ion adsorption by pores in reciprocal space \cite{boukhalfa_2013}, neutron imaging yields a direct transmission profile in real space. This is a non-destructive technique in which the neutrons interact with the nuclei of the sample, in contrast to \mbox{x-rays} which transfer energy to the electrons \cite{kardjilov_2011}. Thus, the element-specificity differs greatly between the methods, a fact that is appreciable when considering the neutron absorption cross section of Gd \mbox{$\sigma_{\mathrm{Gd}}=46\,700$\,barn} for thermal neutrons \cite{mastromarco_2019}, providing unrivalled contrast and excellent properties for the mapping of concentration evolution \cite{sharma_2013,butcher_2020}. Gd also happens to have a large magnetic moment 7\,$\mu_B$, courtesy of the unpaired electrons in its half filled 4f shell.  

Recent advances in detector systems have provided the means for neutron imaging with both high spatial and temporal resolutions \cite{trtik_2016, zboray_2019}. The technique has previously found use in the study of lithium batteries \cite{owejan_2012,song_2019} and capacitive deionization with ordered mesoporous carbon electrodes of ${\sim}10\,$nm pore size \cite{sharma_2013, sharma_2015, kim_2019}. The first of these capacitive deionization studies was restricted to a relatively dilute \Gd solution of 8.74\,mM in a flow-through cell and neutron images were obtained every 5\,min \cite{sharma_2013}. This experiment used cold neutrons, which amplify the absorption cross section of Gd.

Here, the results of \Gd capacitive deionization by carbon aerogels with a broad pore size distribution, spanning both meso- and macropores, are reported. The mesopores are centered at 15\,nm, whereas the maximum of the macropore distribution lies at 115\,nm. The uptake of Gd$^{3+}$ by the carbon aerogel electrodes and its depletion around them were tracked by dynamic neutron imaging. Furthermore, the effect of a magnetic field gradient on the ensuing system, out of equilibrium, was investigated for a possible magnetic modification of the concentration profile.

%----------------------------------------------------------------------------------------------------------------------------------------------------------------------------------------------
\section{Methods}%----------------------------------------------------------------------------------------------------------------------------------------------------------------
%----------------------------------------------------------------------------------------------------------------------------------------------------------------------------------------------

\subsection{Porous Material Characterization}

\begin{figure}
	\centering
	\includegraphics[width=0.85\columnwidth]{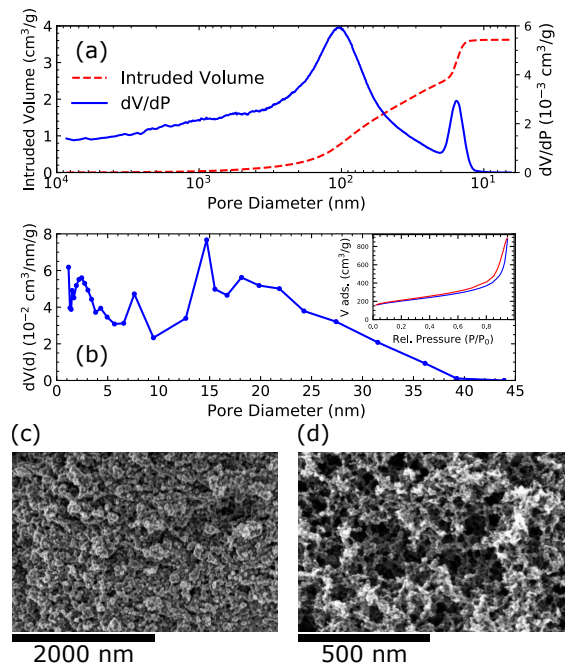}
	\caption{Pore structure of the carbon aerogel sample. (a) Hg porosimetry with two pronounced regions of pores with maxima at 115\,nm and 16\,nm. (b) BJH pore-size distribution of the mesopores; inset: Type IV Nitrogen adsorption isotherm (blue: adsorption, red: desorption). The measured surface area is 720\,m$^2$\,g$^{-1}$.  (c) SEM image of the carbon aerogel. (d) Higher magnification SEM image. The dark voids indicate high porosity. A large portion of the mesopores is below the resolution of the SEM.}
	\label{fig:BET}
\end{figure}

Resorcinol-formaldehyde polymer-derived carbon aerogel monolith disks \cite{pekala_1989, wang_1993} of approximately 12\,mm diameter were purchased from Aerogel Technologies. The disks were concave with 3\,mm thickness on the sides and 2\,mm in the centre. Their bulk density was approximately 0.25\,g\,cm$^{-3}$. A quantitative determination of the pore size distribution was accomplished with mercury (Hg) porosimetry and Brunauer-Emmett-Teller (BET) surface area analysis (see Fig.~\ref{fig:BET}). The instruments used were an Autoscan-33 Porosimeter (Quantachrome, UK) and a Nova 2400e Surface Area Analyser (Quantachrome, UK) with nitrogen gas adsorbate. Mercury porosimetry was performed up to a maximum pressure of 33\,000 psi with a default contact angle of 140°. Prior to analysis, samples were de-gassed for 1\,h at 200\,°C under vacuum.
%(see Fig.~\ref{fig:setup})

Mercury porosimetry revealed pores in the approximate range 7\,nm-10\,\textmu m (see Fig.~\ref{fig:BET}(a)). The dashed curve in Fig.~\ref{fig:BET}(a) shows the intrusion of Hg into the aerogel as a function of pressure (with pressure being analogous to pore diameter). An increase of pressure causes a gradual filling, first of large pores with 2\,\textmu m to 150\,nm diameter. This filling continues until a significant intrusion of mercury occurs for pores $<$150 nm diameter. The intrusion proceeds to approximately 20\,nm, with the total pore volume up to this point equalling 2.5\,cm$^3$\,g$^{-1}$. Then a further pronounced intrusion of mercury occurs in the mesopore range, beyond which the curve plateaus out as all pores are fully filled.
The solid curve in Fig.~\ref{fig:BET}(a) shows the pore size distribution, plotted as the derivative of volume with respect to pressure. Large changes in volume over small pressure ranges yield sharp peaks, and vice versa. A large broad peak with a maximum at 115\,nm occurs in the range 20\,nm-10\,\textmu m. These pores account for 2.5\,cm$^3$\,g$^{-1}$ of the 3.62\,cm$^3$\,g$^{-1}$ total pore volume, which corresponds to 69\%, on a volume basis. The broad peak is flanked by a sharper peak representing a high concentration of mesopores between 10 and 20\,nm. These account for 1.12\,cm$^3$\,g$^{-1}$, or 31\% of the sample, on a volume basis. In order to better understand the mesoporosity, which provides the high surface area for capacitive deionization, BET surface area analysis was carried out. From this analysis, a surface area of 720\,m$^2$\,g$^{-1}$ was measured. The isotherm was type IV, with hysteresis for P/P$_0$ values above approximately 0.6 (see inset of Fig.~\ref{fig:BET}(b)). The Barrett-Joyner-Halenda (BJH) method was used to calculate the pore size diameter and pore volume, from the desorption branch of the isotherm (see Fig.~\ref{fig:BET}(b)). This yielded pores in the range 1-40\,nm and a total adsorbed volume of 1.28\,cm$^3$\,g$^{-1}$. Sub-10\,nm pores account for approximately 0.4\,cm$^3$\,g$^{-1}$ of this. Within this size range lie two peaks: a broad peak at 2.5\,nm and a sharper peak at 7.5\,nm. The remaining mesoporosity is above 10\,nm and accounts for 0.88\,cm$^3$\,g$^{-1}$. The main feature within this size range is the presence of a peak at 14.7\,nm, with further significant porosity occurring up to 40\,nm. The sharp peak at 16.4\,nm in the Hg porosimetry (Fig.~\ref{fig:BET}(a)) is consistent with the maximum at 14.7\,nm observed in the BJH analysis. 

The porous properties agree well with the corresponding morphology, which was imaged with scanning electron microscopy (SEM) (Figs.~\ref{fig:BET}(c) and (d)). The fracture surfaces display widespread porosity. Some clumping and agglomeration exists, which can explain the largest pores detected with porosimetry. The aerogel is composed of a globular network, with individual globules appearing lightly fused and predominantly $<100$\,nm in diameter. The spacing between these gives rise to the primary porosity, circa 100\,nm. This loosely-packed morphology has a high pore volume (2.5\,cm$^3$\,g$^{-1}$), consistent with a low bulk density material having an open, interconnected pore network and 89.1\% porosity.

\subsection{Neutron Imaging Experiment}

Neutron imaging experiments were performed at the NEUTRA station (measuring position 2) \cite{lehmann_2001} in the Paul Scherrer Institute. This station operates with thermal neutrons from a 25\,meV Maxwellian spectrum. The neutron flux at the sample position was approximately $1.3\times 10^7\,\mathrm{cm}^{-2}\, \mathrm{s}^{-1}$. The detection system consisted of a 30\,\textmu m thick terbium doped gadolinium oxysulfide (Gd$_2$O$_2$S:Tb - Gadox: Tb-doped) scintillator, fitted in the MIDI camera box and coupled with a CCD camera (Andor, iKon-L). Recorded images of 2048$\times$2048 pixels corresponded to a field of view of $67.67\,\mathrm{mm}\times67.67$\,mm with a pixel size of 33.04\,\textmu m. Series of neutron images were acquired with an exposure time of 10\,s and read out time of approximately 3\,s. Occasional interruptions of the imaging sequence were caused by fluctuations of the neutron beam intensity.

\begin{figure}
	\centering
	\includegraphics[width=0.8\columnwidth]{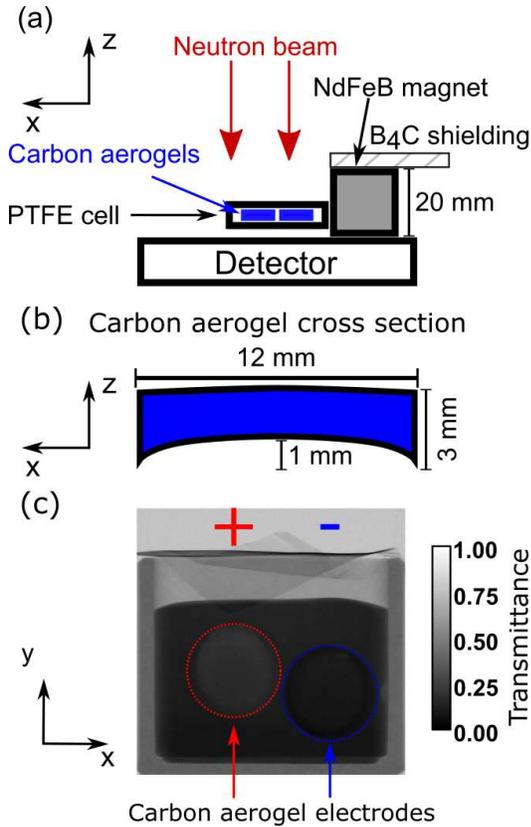}
	\caption{(a) Sketch of the experimental setup (top view). The PTFE cell contained the carbon aerogels and \Gd solution. The aerogels were connected to a potentiostat. (b) Cross section  of the concave carbon aerogel disks. (c) Dark current corrected neutron image (normalized by open beam: $T=\tfrac{I - I_{dc}}{I_0-I_{dc}})$ of the 6\,mm path length PTFE sample holder with 70\,mM \Gd solution: 10\,min into first charge at 1.0\,V. Right aerogel: negative charge (Gd$^{3+}$); Left aerogel: positive charge (NO$_3$$^{-}$). Accumulation of Gd$^{3+}$ ions in the right aerogel manifests itself in lower neutron transmission, whereas the reduced Gd$^{3+}$ concentration leads to a transmittance increase.}
	\label{fig:setup}
\end{figure}

Fibre reinforced teflon (PTFE) cells with outside dimensions of \mbox{34\,mm $\times$ 28\,mm $\times$ 10\,mm} \mbox{(height $\times$ width $\times$ depth)} and a path length of 6\,mm were placed 12\,mm from the detector (see sketch in Fig.~\ref{fig:setup}(a)).

Normal water is a strong incoherent scatterer (\mbox{$\sigma_{\mathrm{H}_2 \mathrm{O}} = 169$\,barn}) that is detrimental to neutron imaging. This was avoided by preparing the \Gd solution in D$_2$O (\mbox{$\sigma_{\mathrm{D}_2 \mathrm{O}} = 20$\,barn}). A molarity of 70\,mM provided the best balance between contrast and concentration with a path length of 6\,mm. This was low enough to avoid the high absorption limit, where the transmittance is dominated by scattering, while retaining the Gd$^{3+}$ concentration for the paramagnetic susceptibility \cite{butcher_2020}. 

Prior to the experiment, the carbon aerogel monoliths were soaked in D$_2$O to fill the pores and avoid the formation of air bubbles during the neutron imaging. An electrical connection to a potentiostat (Biologic SP-300) was maintained by 100\,\textmu m diameter silver wires that were contacted to the aerogels by silver paint. The aerogel electrodes were placed inside the cell, which was then filled with 3\,mL of 70\,mM \Gd solution (see Fig.~\ref{fig:setup}). The aerogels were stabilized in the liquid by buoyancy and their wire connection to the potentiostat. Their concave form can be seen in the cross section in Fig.~\ref{fig:setup}(b). The cell was covered with parafilm to minimize evaporation and exchange with H$_2$O. A DC voltage was  applied to the cell, while changes in the transmission profile were monitored. A dark field corrected neutron image during the first charging process, normalized by the neutron beam with no sample (open beam), is shown in Fig.~\ref{fig:setup}(c).

A 20\,mm \Nd permanent magnet cube was placed next to the cell when the magnetic field gradient force was investigated. The surface of the magnet with a horizontal magnetic field of $B=0.45$\,T was separated from the solution by the 2\,mm thick cell walls. The resulting force distribution was shown in Fig.~\ref{fig:forces}(a). Boron carbide (B$_4$C) shielding protected the magnet from neutron activation by the beam.

The transmitted neutron intensity $I$ by the sample is given by the Beer-Lambert law:

\begin{equation}
	I = I_0 \, \mathrm{exp} \left( -\sum_{i} \sigma_{i} N_A c_{i} \Delta z \right),
\end{equation}

\noindent with the Avogadro constant $N_A$, the concentration $c_i$,  the neutron cross sections $\sigma_{i}$ and the path length $\Delta z$. The argument of the exponential function is summed over all elements in the path of the beam. At high attenuation ($I/I_0<0.2$), the Beer-Lambert law loses its validity due to scattering events and a linear offset must be introduced \cite{butcher_2020}. The choice of 70\,mM \Gd solution precluded this by ensuring $I/I_0>0.25$.   

Variations in the transmitted neutron intensity due to Gd$^{3+}$ concentration changes range from a few percent in the aerogels down to tenths of a percent in the solution. To better visualize these fine changes of Gd$^{3+}$ concentration ($\Delta c_{\mathrm{Gd}}$), the time-sequenced neutron images were normalized by the first image of the series. Then, the Beer-Lambert law was inverted under the assumption that any alteration of the transmitted intensity stemmed exclusively from a movement of Gd$^{3+}$ \cite{sharma_2013}:

\begin{equation}
	\Delta c_{\mathrm{Gd}} = -\mathrm{ln} \left(\frac{I(t_2)/I(t_1)}{\sigma_{\mathrm{Gd}} N_A \Delta z}\right).
\end{equation}

This assumption is reasonable, considering that the neutron cross section of all other constituents in the path of the beam are overshadowed by the absorption cross section of Gd. These are mainly the cross sections of D$_2$O, carbon (\mbox{$\sigma_\mathrm{C}=5.5$\,barn}) and silver (\mbox{$\sigma_\mathrm{Ag}=68.3$\,barn}). This treatment removes any absorption contribution of the immobile components of the sample and scattering is taken into account implicitly by the division of the transmission profiles.

%----------------------------------------------------------------------------------------------------------------------------------------------------------------------------------------------
\section{Results and Discussion}%-------------------------------------------------------------------------------------------------------------------------------------------------------------
%----------------------------------------------------------------------------------------------------------------------------------------------------------------------------------------------

Immediately upon coming into contact with the aerogel, the 70\,mM \Gd solution began to fill the macropores. This process took place in the first 10\,min. The mesopores, however, remain inaccessible to the ions on such fast timescales and require an electric field to force ionic migration into them. A full capacitive deionization cycle was recorded with neutron imaging and is displayed in Fig.~\ref{fig:ca_charge} (see supplementary material for corresponding time-sequenced neutron images). All neutron images were converted to mean values of $\Delta c_\mathrm{Gd}$ along the path length of the cell interior.  The mean concentration change in the aerogels themselves ($\Delta c_\mathrm{Gd}^\mathrm{ae}$) is displayed in  Fig.~\ref{fig:ca_charge}(g), below the neutron images of the corresponding stages of the process. These values were calculated under the premise that $\Delta c_\mathrm{Gd}^\mathrm{ae}$ was entirely due to adsorption in the aerogel within the area defined by their contour.

The capacitive deionization commenced with the application of 1\,V potential difference to the carbon aerogel electrodes immersed in solution. The voltage is accompanied by the movement of electrons from the potentiostat into the porous structure of the aerogel. Ions from the solution form a double layer to compensate the charged surface and a current flows through the solution between the electrodes. The registered movement of Gd$^{3+}$ bears close resemblance to the current distribution shown in Fig.~\ref{fig:forces}(b) with the region of highest electric potential gradient between the disks experiencing an instantaneous change in ion concentration (see Fig.~\ref{fig:ca_charge}(a)). It can be witnessed how Gd$^{3+}$ ions are expelled from the positively charged aerogel disk on the left and migrate into the negatively charged aerogel disk on the right. The double layer is continuously filled with new ions during the charging process. These first originate from the oppositely charged aerogel. Once the co-ion concentrations (ions with the same valence sign as the aerogel electrode) within the respective electrodes are depleted, the aerogels begin to leach ions from the surrounding solution (see Fig.~\ref{fig:ca_charge}(b) and concentration evolution in Fig.~\ref{fig:ca_charge}(g)). This continues until the final capacity has been reached and no further ions can be accommodated at $\Delta c_{\mathrm{Gd}}^\mathrm{ae} \approx 30$\,mM in the negatively charged aerogel. The arrival at this plateau in neutron transmittance was after 60\,min. 

Then, the voltage was switched off and the aerogels discharged (Figs.~\ref{fig:ca_charge}(c) and (d)). The ions trapped in their respective aerogels were liberated and rushed to compensate their corresponding counterions in the opposing electrode. Furthermore, ions diffused out of the pores into the reservoir solution, in which the \Gd concentration increases again (Fig.~\ref{fig:ca_charge}(d)). Inspection of the right aerogel showed that the anodic dissolution of silver as a side reaction took place. This was restricted to the high electric field region on the curved surface between the aerogels. In addition, the formation of Gd$^{3+}$ and NO$_3^-$ pairs in the porous structure was evident. The blocking of mesopores by these ion pairs is an unwanted effect that decreases the efficiency and reversibility of the capacitive deionization process \cite{sharma_2015}. Ion exchange membranes between the electrodes can alleviate this inherent issue by only allowing ions of one charge to pass through to the other side \cite{porada_2013, sharma_2015}.

\begin{figure}
	\centering
	\includegraphics[width=0.98\columnwidth]{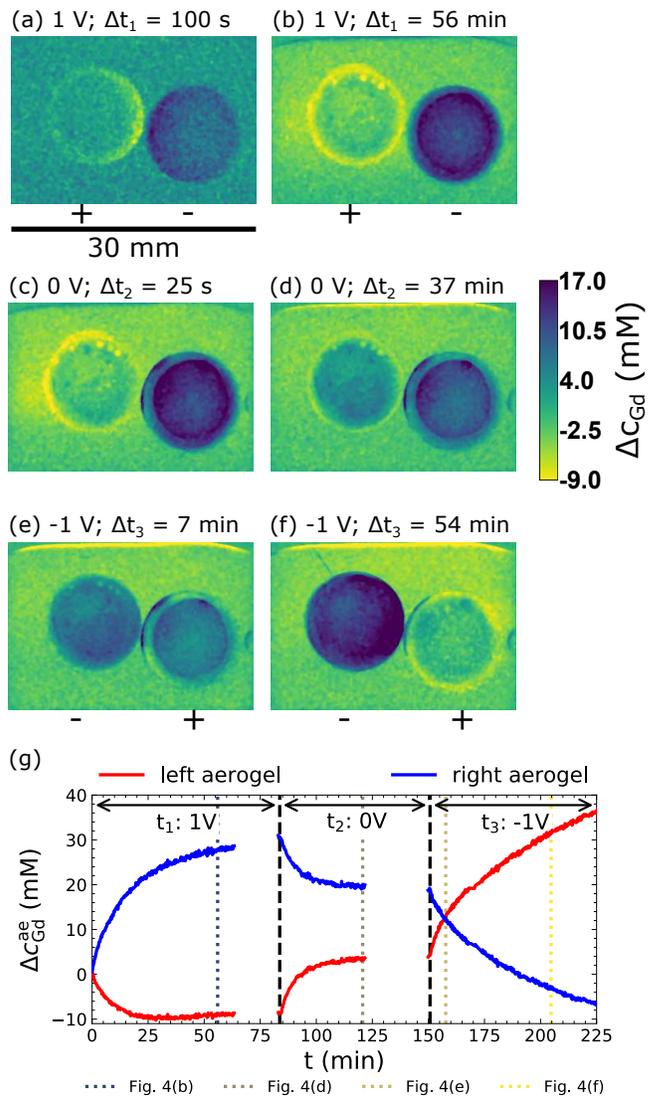}
	\caption{Neutron images converted to $\Delta c_{\mathrm{Gd}}$ during capacitive deionization of 3\,mL 70\,mM \Gd solution by two carbon aerogel disks with 0.5\,mm minimum separation (animations in suppl. material). (a)-(b) Charge: 1\,V, total time $t_1=85$\,min. (c)-(d) Discharge: 0\,V, total time $t_2=65$\,min  (e)-(f) Reverse charge: -1\,V, total time $t_3=82$\,min. Gas bubbles due to oxygen evolution form on the positive electrodes during the charging process. (g) Mean $\Delta c_{\mathrm{Gd}}^\mathrm{ae}$ evolution in the aerogels. During the charge process approximately 10\,mM Gd$^{3+}$ was transferred from the left to the right aerogel. The remaining 20\,mM of Gd$^{3+}$ were adsorbed from the solution. Non-reversibility by the discharge process caused by co-ion adsorption is evident and an inverse voltage is required to unblock the pores.}
	\label{fig:ca_charge}
\end{figure}	

\begin{figure}
	\centering
	\includegraphics[width=0.99\linewidth]{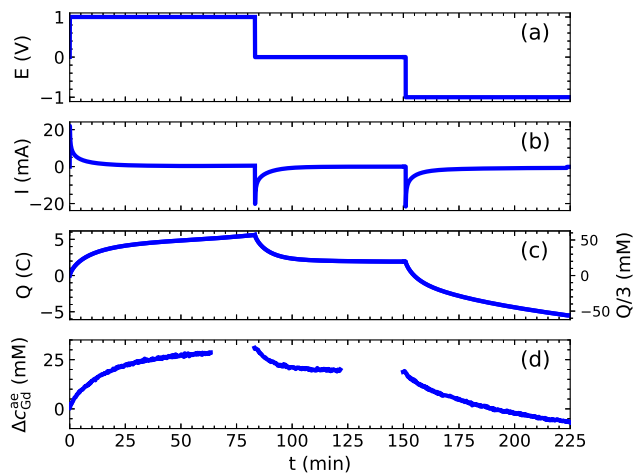}
	\caption{Comparison of the readout from the potentiostat and the mean $\Delta c_{\mathrm{Gd}}^\mathrm{ae}$ in the aerogels detected by neutron imaging. (a-b) Cell voltage and the measured current. (c) Integrated current shows the charge $Q$, which can be converted to a concentration of trivalent ions (see right axis). (d) Neutron imaged $\Delta c_{\mathrm{Gd}}^\mathrm{ae}$ in the right aerogel (see Fig.~\ref{fig:ca_charge}(g)). The charge efficiency $\Lambda_\mathrm{Gd}$ based solely on the capture of Gd$^{3+}$ is around 0.6 during the first charge, but then drops to 0.3.}
	\label{fig:potentiostat}
\end{figure}

After the 65\,min discharge of the electrodes, the aerogels were charged up at the reversed voltage $-1.0$\,V and the Gd$^{3+}$ ions moved in the opposite direction (Figs.~\ref{fig:ca_charge}(e) and \ref{fig:ca_charge}(f)). The dynamics of the electrosorption can be elucidated by inspecting the readout of the potentiostat and comparing it to the $\Delta c_{\mathrm{Gd}}^\mathrm{ae}$ values extracted from the neutron transmission images. The data are summarized in Fig.~\ref{fig:potentiostat}. The upper two panels (Figs.~\ref{fig:potentiostat}(a) and (b)) show the applied cell voltage $E$ and the current response $I$. A value for the transferred charge $Q$ in coulombs is obtained by integration of the current (Fig.~\ref{fig:potentiostat}(c)). This in turn can be converted to a concentration of trivalent ions by division with Faraday's constant and the aerogel volume ($V \approx 0.3$\,mL). Then a comparison with the actually measured $\Delta c_{\mathrm{Gd}}^\mathrm{ae}$ in the right aerogel, which is displayed in the bottom panel in Fig.~\ref{fig:potentiostat}(d) (the blue line from Fig.~\ref{fig:ca_charge}(g)), is possible. The ratio between the adsorbed salt and the measured charge is defined as the charge efficiency $\Lambda$ of the capacitive deionization cell \cite{suss_2015}. Thus, the charge efficiency for Gd$^{3+}$ is calculable by $\Lambda_\mathrm{Gd} = \tfrac{3 \Delta c_\mathrm{Gd}^\mathrm{ae} }{\Delta Q}$, which is approximately 0.6 for the first charge. The value drops to around 0.3 for the second charging processes. Charge efficiencies should approach unity and the calculated values are low.

The main reasons for this have already been mentioned. On the one hand, both the initial expulsion and later adsorption of NO$_3^-$ ions contributes only to the transferred charge, not the neutron imaged $\Delta c_{\mathrm{Gd}}^\mathrm{ae}$. This reaffirms the importance of introducing ion exchange membranes between the electrodes in what is called membrane capacitive deionization \cite{porada_2013}. On the other hand, the anodic dissolution of Ag and subsequent electroplating on the aerogel surface will have taken up significant amounts of the cell current. Thus, the true $\Lambda$ factoring in the charge transfer due to the deposition of Ag is higher than the low $\Lambda_\mathrm{Gd}$ calculated here. In general, electroplating of ions is not necessarily unwelcome during capacitive deionization, as it allows the separation of electroactive ion species. Faradaic reactions at the aerogel surface contribute to parasitic currents. These can range from the reduction of dissolved oxygen to the gradual oxidation of the carbon itself \cite{porada_2013}.

Approximately 75\,min into the second charging process, the right aerogel detached from the wire and sunk to the bottom of the cell. In the wake of this, the system was decidedly out of thermodynamic equilibrium, as the capacitive deionization caused a vertical \Gd concentration gradient. The removal of ions lowered the density of the solution, making it rise due to buoyancy. An estimation of the density change $\Delta \rho$ due to the variation in \Gd concentration is possible with literature values from pycnometric measurements of aqueous rare earth nitrate solutions by Spedding et al. \cite{spedding_1975}.  At concentrations below 1\,M, a linear relationship between $\Delta c_{\mathrm{Gd}}$ and $\Delta \rho$ exists. The coefficient is $\alpha \approx 0.29\,$M$^{-1}$ with respect to the pure solvent density $\rho_0$. Thus, the density change due to  $\Delta c_{\mathrm{Gd}} = 5\,$mM in D$_2$O ($\rho_0 = 1107\,$kg\,m$^{-3}$) is $\Delta \rho = \rho_0 \alpha \Delta c_{\mathrm{Gd}} = 1.605$\,kg\,m$^{-3}$.

\begin{figure}
	\centering
	\includegraphics[width=0.99\columnwidth]{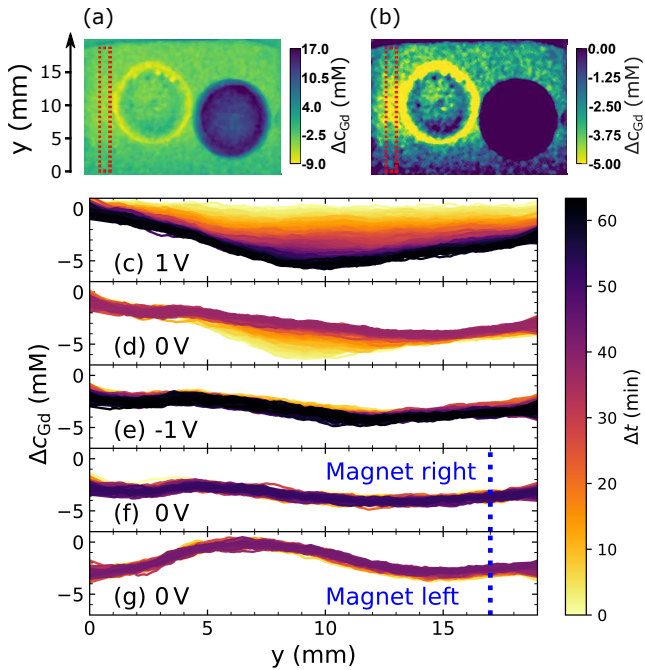}
	\caption{(a) Neutron image converted to $\Delta c_{\mathrm{Gd}}$: 56\,min into charge at 1\,V (70\,mM \Gd solution), see Fig.\,4(b)). (b) Modified contrast for greater visibility of $\Delta c_{\mathrm{Gd}}$ in solution. Vertical profile plots were extracted from the red framed area and are shown in the panels below. (c)-(g) Vertical $\Delta c_{\mathrm{Gd}}$ concentration evolution in liquid left of aerogels \textit{with respect to the initial situation prior to charging} (animations in suppl. material). The individual panels are consecutive and the time displayed by the colour bar is set to restart in each of them. (c) Charge at 1\,V: The liquid is desalinated and a vertical concentration gradient established by the liberated D$_2$O. The desalination is particularly pronounced on the axis of the two aerogels. (d) Discharge at 0\,V: Gd$^{3+}$ leaves the right aerogel and the concentration profile is flattened. The original concentration profile is not re-established. (e) Charge at -1\,V: The concentration profile remains intact, except for a slight decrease in concentration at the bottom of the cell. (f)-(g) 20\,mm \Nd magnet at the side of the cell. The height of the magnet is indicated by the vertical blue dotted lines. (f) Magnet right: The left area is unaffected by the magnet 30\,mm to the right (see Fig.~\ref{fig:magnet_effect}(a)). (g) Magnet left: The magnet attracts 4\,mM of \Gd towards the left side of the cell (see Fig.~\ref{fig:magnet_effect}(b)).}
	\label{fig:v_profiles}
\end{figure}

The emergence of this concentration profile in the solution is shown in Fig.~\ref{fig:v_profiles}. A return to a homogeneous system is imposed by diffusion, which is a lengthy process under normal conditions and even more protracted when considering slow diffusion from the pores of the aerogel. This situation is depicted in the vertical concentration change profiles in the area of the image confined by the dotted lines in Figs.~\ref{fig:v_profiles}(a)-(b). The evolution during the capacitive deionization procedure is shown in Figs.~\ref{fig:v_profiles}(c)-(e). It should be noted that these plots show the progression of $\Delta c_\mathrm{Gd}$ with respect to the initial concentration profile in the cell before the charge process. This was after Gd$^{3+}$ ions had been transported into the macropores via diffusion and convection. A vertical concentration profile was already present in the solution due to this.  

Fig.~\ref{fig:v_profiles}(c) shows $\Delta c_\mathrm{Gd}$ during the first charge at 1\,V. The greatest decline in Gd$^{3+}$ concentration (-6\,mM with respect to the beginning of the charging process) occurred to the left of the aerogels at the height of their horizontal axis. This stratification was mechanically stable, as the initial Gd$^{3+}$ concentration had already decreased by approximately 2\,mM  at the top of the cell. As soon as the electric field was removed, the aerogels released their ions, which flattened the concentration profile. The minimum in $\Delta c_\mathrm{Gd}$ disappeared, but the concentration profile did not return to its original state. Instead, a near-linear concentration profile developed. Switching the voltage to -1\,V did not greatly shift the concentration profile adjacent to the negatively charged aerogel (Fig.~\ref{fig:v_profiles}(e)). One possible reason for this could be that for every trapped trivalent Gd$^{3+}$ ion, three monovalent NO$_3^-$ ions are pulled out of the solution, releasing a greater quantity of desalinated water in a plume around the positively charged aerogel. This can be seen in Figs.~\ref{fig:ca_charge}(b), (c) and (f).

\begin{figure}
	\centering
	\includegraphics[width=0.99\columnwidth]{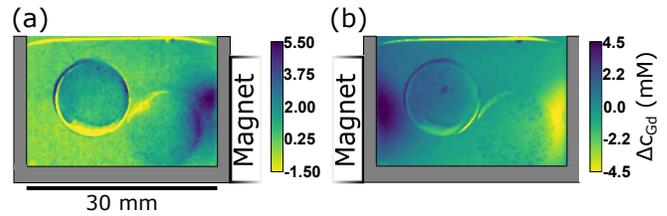}
	\caption{Neutron image converted to $\Delta c_{\mathrm{Gd}}$: Discharging carbon aerogels with 20\,mm magnet cube to sides of the cell. Gd$^{3+}$ ions are drawn into the magnetic field gradient. The right aerogel is disconnected and lies at the bottom of the cell. (a) Magnet to the right: $\Delta c_{\mathrm{Gd}}$ with respect to the end of the desalination process in Fig.~\ref{fig:ca_charge}. (b) Magnet to left: $\Delta c_{\mathrm{Gd}}$ with respect to (a) is shown. The heightened $c_{\mathrm{Gd}}$ has vacated the region on the right and moved to the left with the magnet.}
	\label{fig:magnet_effect}
\end{figure}

Approximately 20\,min after the interruption of the charging process, a 20\,mm Nd-Fe-B cube magnet was placed to the right of the cell to demonstrate a magnetic redistribution of the electrolyte solution (Fig.~\ref{fig:magnet_effect}(a)). The distinctive contour of the magnetic field gradient force (see Fig.~\ref{fig:forces}(a)) is clearly visible as a $\Delta c_{\mathrm{Gd}} \approx 5$\,mM of Gd$^{3+}$ region on the inside of the cell facing the magnet. The concentration profile in the solution on the opposite side of the cell was not influenced by the magnet on the right (Fig.~\ref{fig:v_profiles}(f)). Swapping the position of the magnet after 90\,min from the right to the left of the cell created a symmetric situation with magnetically attracted ions transferring from the right to the left (Fig.~\ref{fig:magnet_effect}(b) and vertical concentration profile in Fig.~\ref{fig:v_profiles}(g)). Evidently, the magnet upset hydrostatic stability and caused movement of the bulk fluid with higher \Gd concentration at the bottom of the cell by convection. This can be understood by comparing the magnetic $E_\mathrm{mag} = \tfrac{\Delta c \chi_m}{2\mu_0} B^2 $ and gravitational $E_\mathrm{grav}=\Delta \rho g \Delta y$ ($g=9.81$\,m\,s$^{-2}$) energy densities. For a concentration gradient of $\Delta c_\mathrm{Gd} = 5\,$mM in a magnetic field of $B=0.35\,$T, these are approximately equal at a height of $\Delta y = 5\,$mm ($E_\mathrm{mag} = E_\mathrm{grav} \approx 80\,$mJ\,m$^{-3}$).

The initial stabilization of the \Gd in the field gradient is almost instantaneous, but as time progresses, the magnetized region could be seen to slightly expand as more ions were drawn into it. Removal of the magnet re-establishes the status quo. The system returns to a homogeneous equilibrium, whether the magnet is present or not. Hence, the concentration profile can only be magnetically manipulated in the time window set by this diffusional process. As a direct consequence, the magnet is also able to refresh the depleted \Gd concentration around a desalinating porous electrode by perpetually dragging magnetic fluid into its vicinity.

%----------------------------------------------------------------------------------------------------------------------------------------------------------------------------------------------

\section{Conclusions}%-------------------------------------------------------------------------------------------------------------------------------------------------------------------------
%----------------------------------------------------------------------------------------------------------------------------------------------------------------------------------------------

Neutron imaging provided a clear record of capacitive deionization of a paramagnetic \Gd solution by disk-shaped carbon aerogel electrodes with a broad pore size distribution comprising meso- and macropores. Both the Gd$^{3+}$ ion transport inside the porous carbon electrode and the diffusion limited desalting of the bulk solution can be captured. A density-difference driven vertical concentration gradient of \Gd was created by the capacitive deionization and a magnetic field gradient was able to disturb hydrostatic stability. The electric driving force from the porous electrodes represents an addition to the previously studied evaporation controlled magnetic enrichment method \cite{eckert_2012, eckert_2014, rodrigues_2017, eckert_2017,  rodrigues_2019, lei_2020, lei_2021}. Specifically, the customisability of the geometry for the harvesting of ions in a cyclical process is noteworthy. Furthermore, activation of the carbon aerogel may improve the capacitive deionization performance by increasing the surface area \cite{hanzawa_1996, yang_2018}. A key point is that ion exchange membranes are necessary to ensure the efficient long-term functioning of the capacitive deionization cell \cite{porada_2013, sharma_2015}.

The magnetic manipulation of the paramagnetic species may prove to be important for the magnetically aided separation of ion species from aqueous solutions. One example of how such a magnetically modified convection front may be exploited is offered by the phenomenon of inverse electrodeposits from solutions containing a non-magnetic electroactive and a paramagnetic non-electroactive ion species \cite{tschulik_2011, tschulik_2012, dunne_2011, dunne_2012}. Here, the curl of the magnetic field gradient causes a flow of the bulk solution away from the magnetized region of the electrode and deposits are thicker in regions of low field. As was demonstrated here, the capacitive deionization process is severely mass-transport limited and a situation where the uptake of ions by the pores is spatially constricted by magnetic field gradients along the electrode is worth examining.

Porous carbon electrodes offer many hitherto unexplored opportunities in combination with magnetic field gradients in general. Research on different ion selective mechanisms in the electrosorbing pores, such as selectivity by hydration or ionic radii, is ongoing \cite{gabelich_2002, gamaethiralalage_2021} and may complement magneto-convection towards the desalinating electrodes. Miniaturization of the electrochemical cell will serve to maximise the concentration and potential gradients within. Such a cell may integrate small magnets or even incorporate a magnetic material directly in the porous electrode \cite{liu_2019,ma_2021}, thus generating larger magnetic field gradients. Previous studies have already demonstrated the accelerated transport of paramagnetic ions through porous membranes in magnetic fields \cite{waskaas_1993, svendsen_2020}. The introduction of a liquid flow \cite{haehnel_2019}, as is routinely done in flow-by capacitive deionization cells, provides ample opportunities for study. 

All this is potentially relevant in the new research area of rare earth ion separation via adsorption in functionalized mesoporous materials \cite{perreault_2017,hu_2018}. Neutron imaging with higher temporal and spatial resolutions \cite{morgano_2015, trtik_2016_1, zboray_2018} along with the unlocking of small-angle neutron scattering information via dark-field imaging \cite{betz_2015, siegwart_2019, valsecchi_2020} offer intriguing possibilities for future investigation of these ideas.

\section*{Supplementary Material}

Time-sequenced images of Gd$^{3+}$ concentration change during capacitive deionization in Fig.~\ref{fig:ca_charge}. The mean values of $\Delta c_{\mathrm{Gd}}$ in the aerogels (Fig.~\ref{fig:ca_charge}(g)) and vertically in the liquid (Figs.~\ref{fig:v_profiles}(c)-(e)) are shown alongside the respective converted neutron images.

%----------------------------------------------------------------------------------------------------------------------------------------------------------------------------------------------
\section*{Acknowledgements}%---------------------------------------------------------------------------------------------------------------------------------------------------------------------
%----------------------------------------------------------------------------------------------------------------------------------------------------------------------------------------------

This work forms part of the MAMI project, which is an Innovative Training Network funded by the European Union’s Horizon 2020 research and innovation program under grant agreement No. 766007. L. P. acknowledges the
support from the Irish Research Council under grant GOIPG/2019/4430. J. M. D. C. is grateful to Science Foundation Ireland for support from contract 16/IA/4534 ZEMS. We thank Niclas Teichert for performing SEM characterization at the CRANN Advanced Microscopy Laboratory and Jong Min Lee for his assistance at the NEUTRA beamline.

\bibliography{neutron_aerogel}

\begin{thebibliography}{65}
\bibitem [{rem()}]{remark_kelvin}%
  \BibitemOpen
  \href@noop {} {}\bibinfo {note} {Note that this expression is only valid if
  $\mathbf{B}\,{\approx}\, \mu_0 \mathbf{H}$.}\BibitemShut {Stop}%
\end{thebibliography}

\end{document}